\documentclass[12pt]{amsart}

\newcommand{\R}{\mathbb R}
\newcommand{\CC}{\mathbb C}
\newcommand{\Imm}{\mathop{\rm Im}\nolimits}
\newcommand{\Sp}{\mathop{\rm Sp}\nolimits}
\newcommand{\Mp}{\mathop{\rm Mp}\nolimits}
\newcommand{\grad}{\mathop{\rm grad}\nolimits}
\newcommand{\Un}{\mathop{\rm U}\nolimits}
\newcommand{\SG}{\mathcal{SG}}
\newcommand{\tr}{\mathop{\rm tr}\nolimits}
\newcommand{\rk}{\mathop{\rm rank}\nolimits}
\newcommand{\const}{\mathop{\rm const}\nolimits}

\begin{document}
\title{Gaussian transform of the Weil representation}

\author{A. V. Stoyanovsky}

\address{Moscow Center for Continuous Mathematical Education}

\thanks{Partially supported by the grant RFFI N~04-01-00640}

\begin{abstract}
A description is given of the image of the Weil representation of the symplectic group
in the Schwartz space and in the space of tempered distributions under the Gaussian integral transform.
We also discuss the problem
of infinite dimensional generalization of the Weil representation in the Schwartz space, in order to construct
appropriate quantization of free scalar field.
\end{abstract}

\email{stoyan@mccme.ru}

\maketitle

\section{Introduction}

This work arose during the study of the problem of quantization of fields, i.~e.,
mathematically and logically self-consistent construction of quantum field theory.
One of formulations of this problem is to give a mathematical sense to the quantum field theory
Schrodinger variational differential equation [3,8] and its relativistically invariant
generalization [13--15].

The Schrodinger equation for free scalar field reads
\begin{equation}
ih\frac{\partial\Psi}{\partial t}=\int\left(-\frac{h^2}2\frac{\delta^2}{\delta u(x)^2}+\frac12(\grad u(x))^2
+\frac{m^2}2u(x)^2\right)\Psi\,dx.
\end{equation}
Here $\Psi$ is an unknown functional depending on a number $t$ and on a real function $u(x)$, $x=(x_1,\ldots,x_k)$;
$\frac{\delta}{\delta u(x)}$ is the variational derivative.

Traditionally one solves equation (1) in the Fock Hilbert space containing functionals of the form
$$
\Psi(u)=\Psi_0(u)\exp\left(-\frac1{2h}\int\hat u(p)\hat u(-p)\omega_pdp\right),
$$
where $\Psi_0$ is a polynomial functional of $u$; $p=(p_1,\ldots,p_k)$,
$\hat u(p)=\frac1{(2\pi)^{k/2}}\int e^{ipx}u(x)dx$, $\omega_p=\sqrt{p^2+m^2}$.
It is easy to see that on these functionals the right hand side of equation (1) equals infinity.
To overcome this, one subtracts an ``infinite constant'' from the Hamiltonian in the right hand side of (1),
and reduces this Hamiltonian to a normally ordered expression of creation operators
$\frac 1{\sqrt{2\omega_p}}\left(-h\frac{\delta}{\delta\hat u(-p)}+\omega_p\hat u(p)\right)$ and annihilation operators
$\frac 1{\sqrt{2\omega_p}}\left(h\frac{\delta}{\delta\hat u(-p)}+\omega_p\hat u(p)\right)$.
This approach meets big problems [8]. One of them is that in the Fock space for $k>1$
one cannot give a mathematical sense
to relativistically invariant generalization of equation (1), i.~e., one cannot perform quantization on spacelike
surfaces~[16].

The idea of our approach to equation (1) was to try to define an analog of the space of main functions and distributions
on the infinite dimensional space of functions
$u(x)$ (let us call them main and distribution functionals),
and to solve equation (1) in these distribution functionals. This approach is also valid from the physical point
of view (unlike the approach of considering dynamics in Hilbert space),
since quantum mechanical quantities like energy and momentum are non-measurable in relativistic quantum dynamics,
and the only measurable quantities are the scattering sections. A negative result of the present paper is that
this approach turned out to be not fruitful
for the problem of giving mathematical sense to equation (1) and quantization on spacelike surfaces (see
Remark 7 at \S6), but it led to some mathematical results which we also present in this paper. The problem
of quantization of a free field on spacelike surfaces is solved in the paper [18].

The space of main functionals that we look for could satisfy the following requirements.

1) It is a locally convex topological vector space with the action of differential operators with polynomial
coefficients, or rather with the action of infinite dimensional analog of the Weyl algebra
(cf., for example, [17], \S18.5).

2) It has an action of infinite dimensional symplectic group. (This is needed, for example, for quantization on
spacelike surfaces. Indeed, in classical field theory the evolution operator of the Klein--Gordon equation
from one spacelike surface to another one is a linear symplectic transformation of the phase space
of a free field.)

3) It has the distribution functional $1$.

4) Its finite dimensional analog is the Weil representation of the symplectic group $\Sp(2n,\R)$ in the Schwartz
space $S(\R^n)$ of rapidly decreasing smooth functions ([5,10]; [17], \S18.5; see \S2 below).

5) It contains Gaussian functionals, i.~e., exponents of a quadratic form.

Let us comment on the last requirement. In the finite dimensional case Gaussian functions are transformed under the
action of the symplectic group in the most simple way. Hence one can expect that in the infinite dimensional case
main functionals will also contain analogs of Gaussian functions.
Besides that, the space of asymptotic states as $t\to\pm\infty$
should be identified with the Fock space, to make the $S$-matrix a unitary operator in this space.
Hence one would like to have an analog of the Gaussian integral, i.~e., Gaussian distribution functionals.

A direct generalization of the Schwartz space to infinite dimensions meets difficulties. Indeed, if, for example,
we define it as a space of weakly smooth functionals on an infinite dimensional (say, nuclear) space,
with the norms of functionals analogous to the norms in the space $S(\R^n)$, then a simplest functional
$$
\Psi(u)=l(u)\exp(-B(u,u))
$$
($l(u)$ is a linear functional, $B(u,u)$ is a positive definite quadratic functional) is in general unbounded,
and hence it does not belong to this space, which contradicts to requirement~5.

Since Gaussian functions are important for us, it is natural to try to consider the finite dimensional case
from the point of view of these functions. This leads to the Gaussian transform, which takes a function
$\psi(x_1,\ldots,x_n)$ to the function
\begin{equation}
u(Z_{jk})=\int\overline{\psi(x_1,\ldots,x_n)}e^{{\frac i2}\sum_{j,k}Z_{jk}x_jx_k}dx_1\ldots dx_n.
\end{equation}
The main purpose of the present paper is to describe the image of the Weil representation under the Gaussian
transform. It turns out that it is rather easy to describe the image of the space
$S(\R^n)$ and of the space $S'(\R^n)$ of tempered distributions, and it is less easy to describe
the image of the space $L_2(\R^n)$ and other spaces, cf. Remarks in \S6.
Thus, we obtain an explicit description of the Weil representation in the spaces
$S(\R^n)$ and $S'(\R^n)$, in which the Gaussian vectors and the vector $1$ play a distinguished role.
This is the main result of the paper.

The paper consists of 6 sections. \S2 contains preliminaries on the Weil representation,
\S3 statements of main theorems, \S\S4,5 proofs, \S6 concluding remarks.

A short exposition of the main theorems of the paper is contained in [19].

The author is grateful to V.~V.~Dolotin and Yu.~A.~Neretin for numerous helpful discussions.

\section{The Weil representation}

\subsection{} The Weil representation arises when one solves the Schrodinger equation with a quadratic Hamiltonian:
\begin{equation}
ih\frac{\partial\psi}{\partial t}=\left(\sum_{j,k}\frac12a_{jk}x_jx_k+ihb_{jk}x_j\frac{\partial}{\partial x_k}
+\frac{h^2}2c_{jk}\frac{\partial}{\partial x_j}\frac{\partial}{\partial x_k}+\frac{ih}2\sum_jb_{jj}\right)\psi.
\end{equation}
Here $\psi(t,x)$ is an unknown function of the variables $t$ and $x=(x_1,\ldots,x_n)$; $a_{jk}$ and $c_{jk}$
are real symmetric matrices; $b_{jk}$ is an arbitrary real matrix.
The summand $\frac{ih}2\sum_jb_{jj}$ is added to make the operator in the right hand side self-adjoint,
and to simplify the formulas below. Put the initial condition
\begin{equation}
\psi(0,x)=\delta(x-y).
\end{equation}
It turns out that exact solution of the problem (3,4) is given by the quasiclassical approximation. It
reads~[6]
\begin{equation}
\psi(t,x)=a(t,h)\exp(iS(t,x,y)/h),
\end{equation}
where $S(t,x,y)$ is the action of the corresponding Hamiltonian system at the time $t$, $a(t,h)$ is the amplitude,
which are computed in the following way. The Hamiltonian flow at the time $t$ gives a linear symplectic transformation
of the phase space $\left(\begin{array}{c}p\\x\end{array}\right)$,
$p=(p_1,\ldots,p_n)^T$. We will denote the points of the phase space by vector columns; the index
$T$ denotes transposing (in the present case, transposing of a row).
Denote the matrix of this transform by $\left(\begin{array}{cc}A&B\\C&D\end{array}\right)\in\Sp(2n,\R)$
($A$,$B$,$C$,$D$ are $n\times n$-matrices),
i.~e.,
$$
\left(\begin{array}{c}q\\y\end{array}\right)\mapsto\left(\begin{array}{c}p\\x\end{array}\right),\
\begin{array}{l}p=Aq+By,\\x=Cq+Dy.\end{array}
$$
Then
\begin{equation}
a(t,h)=1/\sqrt{(2\pi ih)^n\det C},
\end{equation}
and the formula $dS=pdx-qdy-H(x,p)dt$ ($H(x,p)$ is the classical Hamiltonian) implies that
\begin{equation}
S(t,x,y)=\frac12x^TAC^{-1}x-y^TC^{-1}x+\frac12y^TC^{-1}Dy.
\end{equation}
Below we put $h=1$.
Thus, the evolution operator of equation (3) at the time $t$ has the form
\begin{equation}
(U\psi)(x)=\frac1{\sqrt{(2\pi i)^n\det C}}\int e^{i\left(\frac12x^TAC^{-1}x-y^TC^{-1}x+\frac12y^TC^{-1}Dy\right)}\psi(y)dy
\end{equation}
for $\det C\ne 0$. The operator $U$ is unitary and extends to a unitary operator on $L_2(\R^n)$. Indeed, this operator
is the composition of four operators: 1)~the operator of multiplication by the function
$\exp\left(\frac i2y^TC^{-1}Dy\right)$, 2)~the linear change of coordinates with the matrix $(C^T)^{-1}$,
followed by multiplication by $1/\sqrt{\det iC}$, 3)~the Fourier transform,
4)~the operator of multiplication by the function $\exp\left(\frac i2x^TAC^{-1}x\right)$. All these operators
are unitary. They also preserve the spaces $S(\R^n)$ and $S'(\R^n)$. Besides that, they preserve the subspaces
of even and odd functions, which we will denote by the indices respectively
$+$ and $-$, for example, $S'_+$, $(L_2)_-$.

It turns out and it is not difficult to check that the set of operators $U$ given by formula (8) for all matrices
$\left(\begin{array}{cc}A&B\\C&D\end{array}\right)$ with $\det C\ne 0$,
can be extended to a two-valued representation of the group $\Sp(2n,\R)$ in the spaces $(L_2)_\pm$, $S_\pm$, $S'_\pm$,
i.~e., to a representation of a two-fold covering of the group $\Sp(2n,\R)$.
The two-valuedness is related to non-uniqueness of the square root from $(2\pi i)^n\det C$. Let us call this
covering by the {\it metaplectic group}, denoted $\Mp(2n,\R)$. An explicit description of this group
will be given in 2.2.

This representation of the group $\Mp(2n,\R)$ is usually called the {\it Weil representation}.

The Lie algebra of the group $\Sp(2n,\R)$ acts by the differential operators
\begin{equation}
\sum_{j,k}\left(\frac i2a_{jk}x_jx_k+b_{jk}x_j\frac{\partial}{\partial x_k}
+\frac i2c_{jk}\frac{\partial}{\partial x_j}\frac{\partial}{\partial x_k}\right)+\frac12\sum_jb_{jj}.
\end{equation}
These operators are identified with the matrices from the symplectic Lie algebra by commuting with the operators
\begin{equation}
\sum_{j=1}^n\left(v_jx_j+w_ji\frac{\partial}{\partial x_j}\right).
\end{equation}
The operators (10) form a $2n$-dimensional symplectic vector space $V_{2n}$; the symplectic form
on $V_{2n}$ is given by the commutator of operators. This implies (or one checks directly from (8))
that the operator $U$ (8) conjugates operators (10) by the standard action of the matrix
$\left(\begin{array}{cc}A&B\\C&D\end{array}\right)$ on the space $V_{2n}$. In other words,
if we write the coefficients of operator (10) as a column vector $\left(\begin{array}{c}v\\w\end{array}\right)$, then
conjugation by $U$ takes operator (10) to the operator of the form (10) with the coefficients
$\left(\begin{array}{cc}A&B\\C&D\end{array}\right)\left(\begin{array}{c}v\\w\end{array}\right)$.

This property determines the operator $U$ uniquely up to a scalar factor. Indeed,
if $U'$ is another operator with the same property, then the operator $U'U^{-1}$
commutes with the operators $x_j$ and $i\frac{\partial}{\partial x_j}$, and hence it is multiplication by
a constant, as it is not difficult to show.

This implies once more that the operators $U$ form a projective representation of the group $\Sp(2n,\R)$.
In particular, the matrix
$\left(\begin{array}{cc}E&B\\0&E\end{array}\right)$ acts by multiplication by the function
$\exp\left(\frac i2\sum_{j,k}B_{jk}x_jx_k\right)$,
the matrix $\left(\begin{array}{cc}0&E\\-E&0\end{array}\right)$ acts (up to a constant factor)
by Fourier transform,
the matrix $\left(\begin{array}{cc}A&0\\0&(A^T)^{-1}\end{array}\right)$
acts by composition of a linear change of coordinates with the matrix $A$ and multiplication by $\sqrt{\det A}$.
These matrices generate the group $\Sp(2n,\R)$, which gives one more proof of existence of a projective action
of $\Sp(2n,\R)$ with the above described commutation relations with operators (10).

\subsection{Gaussian functions} These are functions
\begin{equation}
\psi_Z(x)=\exp\left(\frac i2\sum_{j,k}Z_{jk}x_jx_k\right),
\end{equation}
where $Z$ is a symmetric complex matrix. We have $\psi_Z\in S'$ iff the imaginary part
$\Imm Z$ is nonnegative definite, $\Imm Z\ge 0$. Further, $\psi_Z\in S$ iff the matrix
$\Imm Z$ is positive definite, $\Imm Z>0$.

The action of the metaplectic group on the Gaussian functions is given by explicit formulas. This is seen
from the fact that the function $\psi_Z$ satisfies the system of equations
$$
\left(i\frac{\partial}{\partial x_k}+\sum_jZ_{jk}x_j\right)\psi_Z=0,\ \ 1\le k\le n.
$$
The action of the matrix $\left(\begin{array}{cc}A&B\\C&D\end{array}\right)$ takes these equations to the equations
\begin{equation}
\sum_j \left(v_{jk}x_j+w_{jk}i\frac{\partial}{\partial x_j}\right)\psi=0,\ \ 1\le k\le n,
\end{equation}
which, for $\det(CZ+D)\ne0$, are equivalent to the equations on the Gaussian function $\psi_{(AZ+B)(CZ+D)^{-1}}$.

The matrices $Z$ with $\Imm Z>0$ form the so called {\it Siegel upper half-plane} ([4],\S50), denoted
$\SG$. This is a homogeneous space of the group $\Sp(2n,\R)$; the matrix
$\left(\begin{array}{cc}A&B\\C&D\end{array}\right)\in\Sp(2n,\R)$ acts by a fractional linear transform:
$$
Z\mapsto(AZ+B)(CZ+D)^{-1}.
$$
The stabilizer of the matrix $Z=iE$ is the unitary group $\Un(n)$.

An explicit calculation shows that
\begin{equation}
U\psi_Z=\frac{1}{\sqrt{\det(CZ+D)}}\psi_{(AZ+B)(CZ+D)^{-1}}.
\end{equation}
Define the {\it metaplectic group} $\Mp(2n,\R)$ as the set of pairs
\begin{equation}
\left(\left(\begin{array}{cc}A&B\\C&D\end{array}\right), \sqrt{\det(CZ+D)}\right),
\end{equation}
where $\left(\begin{array}{cc}A&B\\C&D\end{array}\right)\in\Sp(2n,\R)$, and
$\sqrt{\det(CZ+D)}$ is one of the two continuous branches of the square root from $\det(CZ+D)\ne 0$, $Z\in\SG$.
Then the multiplication law in $\Mp(2n,\R)$ can be defined so that formula
(13) will give a correctly defined single-valued action of this group on the functions $\psi_Z$.

Denote matrices $Z$ with $\Imm Z=0$ by the letter $a$:
\begin{equation}
\psi_a(x)=\exp\left(\frac i2\sum a_{jk}x_jx_k\right).
\end{equation}
The $\Mp(2n,\R)$-orbit of the function $\psi_a$ consists, in general, of distributions.
This orbit can be found as follows.
The action of an element of $\Mp(2n,\R)$ on the function $\psi_a$ takes it to the function satisfying
the system of equations (12) with real coefficients $v_{jk}$, $w_{jk}$.
The operators in the left hand side of the system are linearly independent and pairwise commute, i.~e., form
a basis of a Lagrangian subspace $L$ in $V_{2n}$.
Conversely, for any real Lagrangian subspace $L\subset V_{2n}$ there exists a unique,
up to proportionality, distribution solution $\psi_L$ of the system (12), in which the operators in the left hand side
form a basis in $L$. It is clear that the distribution $\psi_L$ depends only on the subspace $L$
but not on its basis. An example of $\psi_L$ is the delta-function $\delta(x)$ satisfying the system
of equations $x_k\delta(x)=0$, $1\le k\le n$. It is not difficult to see that in the general case
$\psi_L$ is the product of a Gaussian function (15) on certain subspace in $\R^n$ and the delta-function
in the transversal direction. Let us emphasize once more that $\psi_L$ is defined only up to a scalar
factor.

Thus, we have an embedding of the real Lagrangian Grassmannian $\Lambda_n$ into the projectivization of the space $S'$,
which takes $L$ to $\CC\psi_L$. The image of this embedding is the $\Sp(2n,\R)$-orbit of the function $1$.
(Here one can see what is the form of the operator (8) for $\det C=0$: this is an integral operator with
the Gaussian kernel supported on certain subspace of the space $(x,y)$ and corresponding to the Lagrangian
subspace in $\R^{2n}\oplus\R^{2n}$
which is the graph of the symplectic transform $\left(\begin{array}{cc}A&B\\C&D\end{array}\right)$.)
There arises an
$\Mp(2n,\R)$-equivariant complex line bundle $\mu$ on $\Lambda_n$,
whose fiber at the point $L$ is the line $\CC\psi_L$. This bundle is trivialized by the functions $\psi_a$ (15)
on the open dense chart of the Grassmannian parameterized by symmetric matrices $a$. Under change of chart and
trivialization,
given by the action of element (14) of $\Mp(2n,\R)$, the corresponding transition functions can be found
using Maslov's method of canonical operator [6]. Indeed, the function $\psi_L$ corresponds to the Lagrangian subspace
$L$ in the method of canonical operator. Under evolution of the Schrodinger equation (3), the function $\psi_a$ goes to
$$
\psi_{(Aa+B)(Ca+D)^{-1}}\cdot|\det(Ca+D)|^{-1/2}\cdot e^{i\pi k/2}
$$
for some integer $k$ (the Maslov index). The same formula is obtained in another way in the book [17], \S21.6.
For a third way of calculation of transition functions see below.

The orbits of other functions $\psi_Z$ with $\Imm Z\ge 0$ also, in general, consist of distributions. These orbits
form the closure $\overline{\SG}$ of the Siegel upper half-plane in the complex Lagrangian Grassmannian.
For $L\in\overline{\SG}$ the function $\psi_L$ given by corresponding system (12) with complex coefficients,
is in general the product of the Gaussian function $\psi_Z$ (11) on certain subspace in $\R^n$, with $\Imm Z\ge 0$,
and the delta-function in the transversal direction. $\Sp(2n,\R)$-orbits are parameterized by the rank
of the matrix $\Imm Z$ on the subspace in $\R^n$ which is the support of the function $\psi_L$.

The line bundle $\mu$ extends to an $\Mp(2n,\R)$-equivariant line bundle $\mu_1$ on the closure
$\overline{\SG}$, trivialized by the functions $\psi_Z$ over $\SG$. This and formula (13) imply
a formula for the transition functions of the bundle $\mu$: the action of the element (14) corresponds
to the transition function
\begin{equation}
\lim\limits_{\Imm Z\to 0}1/\sqrt{\det(CZ+D)}.
\end{equation}
(Cf. with the method of complex germ [7,8].)

\subsection{Gaussian transform} For a function $\psi(x)$ its Gaussian transform
$u(Z)=u_\psi(Z)$ is given by formula (2). Under the action of the group $\Mp(2n,\R)$ on the function $\psi(x)$
its Gaussian transform $u_\psi(Z)$ transforms by an explicit formula, see Theorem~5 below.
The main purpose of this paper is to describe the image of the Weil representation
in the space $S$ under the Gaussian transform.

First let us find what obvious necessary conditions the function $u_\psi(Z)$ satisfies for $\psi\in S$.
The function $u_\psi(Z)$ is well defined as a holomorphic function on the Siegel upper half-plane, which satisfies
the following system of partial differential equations:
\begin{equation}
\frac{\partial}{\partial Z_{jl}}\frac{\partial}{\partial Z_{km}}u=
\frac{\partial}{\partial Z_{jm}}\frac{\partial}{\partial Z_{kl}}u,
\end{equation}
where the operators $\frac{\partial}{\partial Z_{jk}}$, for any $j,k$, are defined from the equalities
\begin{equation}
du=\sum_{j,k}\frac{\partial}{\partial Z_{jk}}u\ dZ_{jk}
=\sum_{j<k}2\frac{\partial}{\partial Z_{jk}}u\ dZ_{jk}+\sum_j\frac{\partial}{\partial Z_{jj}}u\ dZ_{jj}.
\end{equation}
This system and the holomorphness condition are obtained by differentiation under the sign of the integral (2).
For an odd function $\psi\in S_-$, the function $u_\psi(Z)$ vanishes.

Further, the function $u$ extends on the boundary of the Siegel upper half-plane as the section
\begin{equation}
u(L)=u_\psi(L)=(\psi,\psi_L)
\end{equation}
of the line bundle $\mu_1^*$ dual to $\mu_1$.
This section is continuous together with all its derivatives with respect
to the action of the Lie algebra of $\Sp(2n,\R)$. In particular, one defines an infinitely differentiable function
\begin{equation}
u(a)=u_\psi(a)=\int\overline{\psi(x)}e^{\frac i2\sum a_{jk}x_jx_k}dx,
\end{equation}
satisfying the system of equations
\begin{equation}
\frac{\partial}{\partial a_{jl}}\frac{\partial}{\partial a_{km}}u=
\frac{\partial}{\partial a_{jm}}\frac{\partial}{\partial a_{kl}}u.
\end{equation}
It will be shown below that this system is invariant under the change of a chart and a trivialization
by action of an element of $\Mp(2n,\R)$.

Let us also define an analog of the Gaussian transform for odd functions $\psi(x)\in S_-$.
This transform is given by the formula
\begin{equation}
u_l(Z)=u_{l,\psi}(Z)=\int\overline{\psi(x)}x_le^{\frac i2\sum Z_{jk}x_jx_k}dx,\ \ 1\le l\le n.
\end{equation}
The holomorphic vector valued function $(u_{l,\psi}(Z))$ on the Siegel upper half-plane satisfies
the system of equations
\begin{equation}
\frac{\partial}{\partial Z_{jk}}u_l(Z)=\frac{\partial}{\partial Z_{jl}}u_k(Z).
\end{equation}
The vector valued function $(u_{l,\psi}(Z))$ extends to a section
\begin{equation}
u_{v,w}(L)=u_{v,w,\psi}(L)=\left(\psi,\sum_{j=1}^n\left(v_jx_j+w_ji\frac{\partial}{\partial x_j}\right)\psi_L\right)
\end{equation}
of the vector bundle $W\otimes\mu_1^*$ on the closure of the Siegel upper half-plane. Here $W$ is the tautological
$n$-dimensional vector bundle over $\overline{\SG}$, whose fiber over a point $L\in\overline{\SG}$ is the space
$L$ itself or the isomorphic space $(V_{2n}/L)^*$.
The right hand side of (24) gives a linear functional on the space $V_{2n}$ of vectors
$(v,w)$, vanishing on the subspace $L$, i.~e., an element of $(V_{2n}/L)^*$.

The section $u_{v,w,\psi}(L)$ is also continuous together with all derivatives with respect to the action of the Lie
algebra of $\Sp(2n,\R)$. In particular, one defines an infinitely differentiable vector valued function
\begin{equation}
u_l(a)=u_{l,\psi}(a)=\int\overline{\psi(x)}x_le^{\frac i2\sum a_{jk}x_jx_k}dx,\ \ 1\le l\le n,
\end{equation}
satisfying the system of equations
\begin{equation}
\frac{\partial}{\partial a_{jk}}u_l(a)=\frac{\partial}{\partial a_{jl}}u_k(a).
\end{equation}
It will be shown below that this system is also invariant under change of a chart and a trivialization
by action of an element of $\Mp(2n,\R)$.

\section{Theorems}

{\it The main theorem for even functions.}

{\bf Theorem 1.} The transform (20) identifies the space $S_+$ with the space of smooth
functions $u(a)$ satisfying equations (21) and extending to sections of the bundle
$\mu_1^*$ on the closure of the Siegel upper half-plane,
which are holomorphic on the upper half-plane and continuous together with all derivatives with respect to the action
of the Lie algebra of $\Sp(2n,\R)$.
The corresponding holomorphic function $u(Z)$ on the Siegel upper half-plane
automatically satisfies equations (17).
\medskip

The analog for odd functions:
\medskip

{\bf Theorem 2.} The transform (25) identifies the space $S_-$ with the space of smooth vector valued functions
$(u_l(a))$  satisfying equations (26) and extending to sections
of the bundle $W\otimes\mu_1^*$ on the closure of the Siegel upper half-plane,
which are holomorphic on the upper half-plane
and continuous together with all derivatives with respect to the action of the Lie algebra of $\Sp(2n,\R)$.
The corresponding holomorphic vector valued function $(u_l(Z))$ on the Siegel upper half-plane
automatically satisfies equations (23).
\medskip

The following theorems are used in the proof of Theorems 1 and 2. They are more easy to prove, because they
deal with distributions.

The following theorem was stated as a conjecture by Yu.~A.~Neretin.
\medskip

{\bf Theorem 3.} The transform (2) identifies the space $S'_+$ with the space of holomorphic functions $u(Z)$
on the Siegel upper half-plane which satisfy equations (17) and have polynomial growth near the boundary of the half-plane,
more precisely, which satisfy the estimate
\begin{equation}
|u(Z)|\le C(1+|Z|)^M(1+|\det(\Imm Z)|^{-1})^N
\end{equation}
for some $C,M,N$. Here $|Z|$ is any norm on the space of matrices.
\medskip

The analog for odd distributions:
\medskip

{\bf Theorem 4.} The transform (22)
identifies the space $S'_-$ with the space of holomorphic vector valued functions $(u_l(Z))$ on $\SG$,
whose components satisfy equations (23) and the estimate (27).
\medskip

{\bf Theorem 5.} Under the identification from Theorem 3, the element inverse to the element (14) of $\Mp(2n,\R)$
takes the function $u(Z)$ to the function
\begin{equation}
v(Z)=\frac1{\sqrt{\det(CZ+D)}}u((AZ+B)(CZ+D)^{-1}).
\end{equation}
Under the identification from Theorem 4 the same element takes the vector valued function $(u_l(Z))$ to the function
$(v_l(Z))$, where
\begin{equation}
\sum_l((CZ+D)^T)_k^lv_l(Z)=\frac1{\sqrt{\det(CZ+D)}}u_k((AZ+B)(CZ+D)^{-1}).
\end{equation}
\medskip

\section{Proofs of Theorems 3,4,5}

\subsection{Proof of Theorem 3}

To make the Gaussian transform of a distribution $\psi(x)$, one should first construct its direct image under the map
\begin{equation}
x\mapsto b,\ \  b=(b_{jk}),\ \  b_{jk}=x_jx_k,
\end{equation}
and then apply the Fourier--Laplace transform to the result.
To prove Theorem 3, one should invert these transforms. For this it almost suffices to use more or less standard
facts on distributions, exposed, for example, in [17], vol.~1.

Let us first prove that for $\psi\in S'_+$ the function $u_\psi(Z)$ satisfies the estimate (27). For that let us
represent the distribution $\psi$ in the form
$\frac{\partial^\alpha}{\partial x^\alpha}\psi_1(x)$,
where $\psi_1(x)$ is a continuous function of polynomial growth, $\alpha$ is a multiindex.
It suffices to integrate (2) by parts, to eliminate $\frac{\partial^\alpha}{\partial x^\alpha}$, and estimate
the obtained Gaussian integral, using the formula
$$
\int ix_jx_k\psi_Z(x)dx=\frac{\partial}{\partial Z_{jk}}\int\psi_Z(x)dx.
$$

Let us now assume that the function $u(Z)$ satisfies the conditions of the theorem, and let us find a function
$\psi\in S'_+$ such that $u=u_\psi$.

Firstly, it is not difficult to show that the function $u(Z)$ has a distribution boundary value --- the limit $u_0(a)$
of functions $u(a+iy)$ as functions of $a$ as $y\to0$, $y>0$ in the space
$S'(\R^{\frac{n(n+1)}2})$ of tempered distributions on the space $\R^{\frac{n(n+1)}2}$
of real symmetric matrices $a$.

Further, consider the inverse Fourier transform of the distribution $u_0$. This is a distribution
$\hat u_0\in S'(\R^{\frac{n(n+1)}2})$ on the space $\R^{\frac{n(n+1)}2}$
of real symmetric $n\times n$-matrices $b$.
The pairing between $a$ and $b$ is given by the formula
$(a,b)=\frac12 \tr ab$. It is not difficult to see that the function $u(Z)$ is the Fourier--Laplace
transform of the distribution
$\hat u_0$ and that $\hat u_0$ is supported on the cone $\{b\ge0\}$
of nonnegative definite matrices, dual to the cone $\{y\ge0\}$.
By equations (17) the distribution $\hat u_0$ satisfies the equations
\begin{equation}
(b_{jl}b_{km}-b_{jm}b_{kl})\hat u_0=0;
\end{equation}
in particular, it is supported on the cone of matrices of rank $1$.

Now Theorem 3 follows from the following Lemma.
\medskip

{\bf Lemma 1.} The direct image under the map (30)
identifies the space $S'_+(\R^n)$ with the space of distributions $\hat u_0\in S'(\R^{\frac{n(n+1)}2})$
supported on the cone $\{b\ge 0, \rk b=1\}$ and satisfying equations (31).
\medskip

Proof of Lemma 1 is based on the Borel theorem on construction of a smooth function with a given Taylor series at $0$.
This proof is elementary, and we omit it for shortness.
\medskip

Theorem 3 is proved.

\subsection{Proof of Theorem 4}

It is not difficult to deduce Theorem 4 from Theorem 3.
Functions $u_l(Z)$ satisfying conditions of the theorem,
also satisfy equations (17). By Theorem 3 one finds distributions $\psi_l\in S'_+(\R^n)$ such that
$u_l=u_{\psi_l}$. We deduce from equations (23) that $x_k\psi_l=x_l\psi_k$ for all $k,l$.
This implies that $\psi_l=x_l\psi$ for some $\psi\in S'_-$. This $\psi$ is the required one, q.~e.~d.

\subsection{Proof of Theorem 5}

For $u(Z)=u_\psi(Z)$, $\psi\in S'_+(\R^n)$, and for the element (14), denoted shortly by
$\left(\begin{array}{cc}A&B\\C&D\end{array}\right)$, we have
$$
\begin{aligned}{}
v(Z)&=\left(\left(\begin{array}{cc}A&B\\C&D\end{array}\right)^{-1}\psi,\psi_Z\right)
=\left(\psi,\left(\begin{array}{cc}A&B\\C&D\end{array}\right)\psi_Z\right)\\
&=\left(\psi,\frac1{\sqrt{\det(CZ+D)}}\psi_{(AZ+B)(CZ+D)^{-1}}\right)\\
&=\frac1{\sqrt{\det(CZ+D)}}u((AZ+B)(CZ+D)^{-1})
\end{aligned}
$$
by formula (13). Similarly one deduces formula (29), but for computation of
$\left(\begin{array}{cc}A&B\\C&D\end{array}\right)x_l\psi_Z$ one should use the commutation relation
between the element $\left(\begin{array}{cc}A&B\\C&D\end{array}\right)$ and the operator $x_l$. Q.~E.~D.
\medskip

\subsection{Remark} A direct computation shows that equations (17), (23) are invariant with respect to the actions
(28), (29). Hence we obtain passing to the limit (16) that equations (21), (26) are invariant
on the Grassmannian $\Lambda_n$.

\section{Proofs of Theorems 1,2}

\subsection{Proof of Theorem 1}

The main idea of proof is the following. First, for a function $\psi\in S_+$ we express its $L_2$-norm
through the function $u_\psi(a)$, and then for arbitrary function $u(a)$ satisfying conditions of the theorem,
we prove that this expression is finite.
This will imply that $u=u_\psi$ for some $\psi\in L_2$.
Similarly we prove that $x^\alpha\frac{\partial^\beta}{\partial x^\beta}\psi\in L_2$ for any multiindices
$\alpha$, $\beta$.

{\bf Step 1.} Let $\psi\in S_+$. Let us express the $L_2$-norm of $\psi$ through the function
$u(a)=u_\psi(a)$.

To this end, let us use the following argument related to the Fourier transform of a density on a submanifold in
$\R^N$ and taken from the proof of Theorem 7.1.26 in [17]. Denote $b_{jk}=x_jx_k$ for $j<k$,
$b_{jj}=x_j^2/2$. Let us make the change of variables $x_j\to b_{1j}$. We obtain
\begin{equation}
\begin{aligned}{}
&\int|\psi(x)|^2dx=2\int\limits_{x_1>0}|\psi(x)|^2dx\\
&=\const\cdot\int\limits_{b_{11}>0}\frac{|\psi_1(b_{11},\ldots,b_{1n})|^2}{b_{11}^{n/2}}db_{11}\ldots db_{1n},
\end{aligned}
\end{equation}
where $\psi_1(b_{11},\ldots,b_{1n})=\psi(x)$.
Further, for $b_{11}>0$, $k>j\ge 2$ we have $b_{jk}=b_{1j}b_{1k}/2b_{11}$, $b_{jj}=b_{1j}^2/4b_{11}$, and
$$
\begin{aligned}{}
&u(a)=\int\overline{\psi(x)}e^{i\sum\limits_{k\ge j}a_{jk}b_{jk}}dx\\
&=\const\cdot\int\limits_{b_{11}>0}\overline{\psi_1(b_{11},\ldots,b_{1n})}
e^{i\sum\limits_{k\ge j}a_{jk}b_{jk}}\frac{db_{11}\ldots db_{1n}}{b_{11}^{n/2}},
\end{aligned}
$$
whence
\begin{equation}
\const\cdot\tilde u(b_{1j})=\left\{\begin{array}{ll}
\overline{\psi_1(b_{11},\ldots,b_{1n})}
e^{i\sum\limits_{k\ge j\ge 2}a_{jk}b_{jk}}/b_{11}^{n/2},& b_{11}>0,\\
0,& b_{11}\le 0.
\end{array}
\right.
\end{equation}
Here $\tilde u(b_{1j})$ is the inverse Fourier transform of the function $u(a)$,
considered as a distribution in $a_{11}$,
$a_{12},\ldots,a_{1n}$ with fixed $a_{jk}$ for $j,k\ge 2$. Let us fix, once and for all, $a_{jk}=-\delta_{jk}$
for $j,k\ge 2$, and denote $a_{11}=s$, $a_{12}=t_2$, $a_{13}=t_3$, $\ldots$, $a_{1n}=t_n$,
$u(a)=u(s,t)$, $t=(t_2,\ldots,t_n)$. Then
\begin{equation}
\const((s+i0)^{-n/2-1}*u(s,t))\tilde{\ \ }=\left\{\begin{array}{ll}
\overline{\psi_1(b_{11},\ldots,b_{1n})}
e^{i\sum\limits_{k\ge j\ge 2}a_{jk}b_{jk}},& b_{11}>0,\\
0,& b_{11}\le 0.
\end{array}
\right.
\end{equation}
Here $*$ denotes convolution with respect to the variable $s$; $(s+i0)^{-n/2-1}$ is proportional to the Fourier
transform of the function $(b_{11})_+^{n/2}$. Formulas (33) and (34) imply,
by the Parcevale equality, that the $L_2$-norm (32)
of $\psi$ equals the integral
\begin{equation}
\const\cdot\int\overline{u(s,t)}((s+i0)^{-n/2-1}*u(s,t))dsdt.
\end{equation}

{\bf Step 2.} Let us prove that for an arbitrary function $u(a)$ satisfying conditions of the theorem, the integral (35)
is absolutely convergent and nonnegative.

To this end, let us use the conditions at infinity following from the fact that
$u(a)$ extends to a smooth section of the bundle $\mu^*$ on the Grassmannian $\Lambda_n$.
Let us compute the change of trivialization of the section $u$ by formula (16) for two matrices
$\left(\begin{array}{cc}A&B\\C&D\end{array}\right)$: 1)~the matrix corresponding to the Fourier transform
in the variable $x_1$,
2)~the matrix corresponding to the Fourier transform in all variables $x_1,\ldots,x_n$. The first one gives the formula
\begin{equation}
u(s,t)=s^{-1/2}v(s^{-1},t_js^{-1},t_jt_ks^{-1}),
\end{equation}
and the second one gives the formula
\begin{equation}
u(s,t)=(s+r^2)^{-1/2}w((s+r^2)^{-1},t_j(s+r^2)^{-1},t_jt_k(s+r^2)^{-1}),
\end{equation}
where $r=\sqrt{t_2^2+t_3^2+\ldots+t_n^2}$; $v$ and $w$ are smooth functions defined for all values of their
arguments, including the zero values.

Let us divide the integration domain of the integral (35) into three subdomains.

1) $|s|\le2$, $r\le2$. This domain is bounded, and the integral is finite in it.

2) $s<-2$, $r^2<|s|^{1+\varepsilon}$, where $\varepsilon>0$ is sufficiently small (less than $1/n$).

In this domain we have
\begin{equation}
|u(s,t)|\le\const|s|^{-1/2}
\end{equation}
by formula (36), since the arguments of the function $v$ are bounded in this domain. Further, we have
\begin{equation}
\begin{aligned}{}
&|(s+i0)^{-n/2-1}*u(s,t)|\le\const\int\limits_{-1}^1|s-s_1|^{-n/2-1}ds_1\\
&+\const\left|\left(\int\limits_1^\infty
 +\int\limits_{-\infty}^{-1}\right)(s-s_1+i0)^{-n/2-1}s_1^{-1/2}v(s_1^{-1})ds_1\right|\\
&\le\const|s|^{-n/2-1}\left(1+\left|\int\limits_{-1}^0(s_2-s^{-1}+i0)^{-n/2-1}s_2^{\frac{n-1}2}v(s_2)ds_2\right|\right)\\
&\le\const|s|^{-1/2-n/2}.
\end{aligned}
\end{equation}
The latter inequality is obtained by integration by parts $m$ times, where $m=\frac n2$ for $n$ even and
$m=\frac{n+1}2$ for $n$ odd, and then, in the case of $n$ even, by dividing the integral into the sum of three
estimated integrals: from $-1$ to $\frac32s^{-1}$, from $\frac32s^{-1}$ to $\frac12s^{-1}$, and from $\frac12s^{-1}$ to $0$.

Now the absolute value of the integral (35) over this domain is less than
$$
\const\cdot\int\limits_{\begin{subarray}{c}s<-2\\r^2<|s|^{1+\varepsilon}\end{subarray}}|s|^{-1-n/2}r^{n-2}drds
=\const\cdot\int\limits_{s<-2}|s|^{-\frac32+\frac{n-1}2\varepsilon}ds<\infty.
$$

3) $|s|\le2$, $r\ge2$ or $s>2$ or $s<-2$, $r^2>|s|^{1+\varepsilon}$. In this domain, similarly to the estimates
(38), (39), from formula (37) we deduce the estimates
$$
\begin{aligned}{}
|u(s,t)|&\le\const(s+r^2)^{-1/2},\\
|(s+i0)^{-n/2-1}*u(s,t)|&\le\const(s+r^2)^{-1/2-n/2},
\end{aligned}
$$
and the absolute value of the integral over this domain is less than
$$
\begin{aligned}{}
&\const\left(\int\limits_{\begin{subarray}{c}|s|\le2\\r\ge2\end{subarray}}+\int\limits_{s>2}
+\int\limits_{\begin{subarray}{c}s<-2\\r^2>|s|^{1+\varepsilon}\end{subarray}}\right)(s+r^2)^{-1-n/2}r^{n-2}drds\\
&\le\const\left(\int\limits_2^\infty r^{-4}dr+1+\int\limits_1^\infty r^{-2}dr\right.\\
&\left.+\int\limits_{2^{\frac{1+\varepsilon}2}}^\infty\left((r^2-2)^{-n/2}-\left(r^2-r^{\frac2{1+\varepsilon}}\right)^{-n/2}
\right)r^{n-2}dr\right)<\infty.
\end{aligned}
$$

Nonnegativity of the integral (35) follows from the fact that it can be presented in the form
$$
\const\cdot\int|(s+i0)^{-n/4-1}*u(s,t)|^2dsdt,
$$
at least for a function $u(a)$ with compact support. The general case is reduced to this one by multiplication
of the function $u(a)$ by a smooth function $0\le\rho(a)\le 1$ with compact support, equal to $1$
in a large neighborhood of the origin.

{\bf Step 3.} Let $u(a)$ be a function satisfying the conditions of the theorem. Let us prove that $u=u_\psi$
for some $\psi\in S'_+$.

Due to formula (16) for the transition function of the bundle $\mu$, the function $u(a)$ is bounded and, moreover,
it tends to zero as $O(|\det(Ca+D)|^{-1/2})$ as $a\to\infty$
in any direction, for appropriate matrices $C$ and $D$. Therefore, the function $u(Z)$ is also bounded,
because the closure of each
$\Sp(2n,\R)$-orbit is compact, and one can use induction with respect to the dimension of an orbit.
Hence $u(Z)$ coincides with the Fourier--Laplace transform of $\hat u$ (see proof of Theorem~3).
Hence the extension $u(Z)$ is unique. Besides that, $u(Z)$ satisfies equations (17).
Indeed, the difference of the left and right hand sides of these equations is defined as a section of the bundle
$\mu_1^*$ on $\overline{\SG}$ (because the operator $\frac{\partial}{\partial Z_{jk}}$
belongs to the Lie algebra of $\Sp(2n,\R)$) and vanishes on the Grassmannian $\Lambda_n$.
Hence by Theorem 3 there exists a unique distribution $\psi\in S'_+$ such that $u=u_\psi$.
It remains to prove that $\psi\in S_+$.

{\bf Step 4.} Let us prove that $\psi\in L_2$.

Denote the Hermitian inner product on the space of functions $u$ satisfying the conditions of the theorem,
given by the integral (35), by $\langle\ ,\rangle$. Consider a linear functional on the space
of functions $\varphi\in S_+$ given by the formula
\begin{equation}
\varphi\mapsto\langle u_\psi,u_\varphi\rangle.
\end{equation}
This functional can be expressed through $\psi$ and $\varphi$ in the following way.
Assume that the function $\varphi(x)$ has compact support and vanishes in a neighborhood
$|x_1|<\varepsilon$ of the hyperplane $x_1=0$. Then
\begin{equation}
\langle u_\psi,u_\varphi\rangle=(\psi,\varphi).
\end{equation}
Indeed, in this case the function $\varphi$
can be expressed as a smooth function $\varphi_1(b_{1j})$ of $b_{11},\ldots,b_{1n}$,
vanishing for $b_{11}<\varepsilon^2/2$.
The number $(\psi,\varphi)$ equals the pairing $(\psi_1,\varphi_1)$ with the direct image $\psi_1(b)$
of the distribution $\psi$ to the space of variables $b_{jk}$. This pairing is well defined since
the intersection of supports of the functions $\psi_1(b)$ and $\varphi_1(b)$ is compact.
It also equals to the pairing $(\psi_2,\varphi_2)$, where
$$
\begin{aligned}{}
\psi_2(b)&=\psi_1(b)\exp\left(-i\sum_{k\ge j}a^0_{jk}b_{jk}\right),\\
\varphi_2(b_{1j})&=\varphi_1(b_{1j})\exp\left(-i\sum_{k\ge j\ge 2}a^0_{jk}b^0_{jk}\right),
\end{aligned}
$$
and $a^0_{1k}=a^0_{k1}=0$, $a^0_{jk}=-\delta_{jk}$ for $j,k\ge 2$, as fixed above;
$b^0_{jk}=b_{1j}b_{1k}/2b_{11}$ for $k>j$, $b^0_{jj}=b_{1j}^2/4b_{11}$, as above.

This pairing is equal to the pairing $(\varphi_3,\psi_3)$, where
$\psi_3(a)$ (respectively $\varphi_3(a)$) is the Fourier transform
of the distribution $\overline{\psi_2(b)}$ (respectively $\overline{\varphi_2(b)}$).
We have
$$
\begin{aligned}{}
\psi_3(a)&=u_\psi(a+a^0),\\
\varphi_3(a)&=\const((s+i0)^{-n/2-1}*u_\varphi(s,t))\cdot\prod_{k\ge j\ge 2}\delta(a_{jk})
\end{aligned}
$$
(see (34)). This implies (41).

Furthermore, by the Cauchy inequality for the scalar product $\langle\ ,\rangle$, the functional (40)
is continuous with respect to the $L_2$-norm of the function $\varphi$. This implies that $\psi$ equals to the sum
of an $L_2$-function and a distribution supported at the hyperplane $x_1=0$.

Applying the same argument to other variables $x_j$ instead of $x_1$ and to other columns of the matrix $b_{jk}$,
we obtain that $\psi$ equals to the sum of an $L_2$-function and a distribution supported at the point $x=0$.

Now consider the Fourier transform $\hat\psi$ of the function $\psi$. It equals to the sum of an $L_2$-function
and a polynomial.
On the other hand, the Fourier transform (up to a constant factor)
belongs to the action of the metaplectic group, and this action preserves the space of functions
$u$ satisfying the conditions of the theorem.
Applying the previous argument to the function
$u_{\hat\psi}$ instead of $u$, we obtain that
$\hat\psi$ is also the sum of an $L_2$-function and a distribution supported at the point $x=0$.
Therefore, $\psi\in L_2$, as required.

{\bf Step 5.} Now let us apply the previous argument to the distributions
$x^\alpha\frac{\partial^\beta}{\partial x^\beta}\psi$. Since the operators $ix_jx_k$,
$x_j\frac{\partial}{\partial x_k}+\frac12\delta_{jk}$,
and $i\frac{\partial}{\partial x_j}\frac{\partial}{\partial x_k}$ belong to the Lie algebra of $\Sp(2n,\R)$ (see~(9)),
and this action preserves the space of functions $u$ satisfying conditions of the theorem,
we obtain that the distribution
$x^\alpha\frac{\partial^\beta}{\partial x^\beta}\psi$ belongs to $L_2$ for any $\alpha$, $\beta$
with even $|\alpha|+|\beta|$. From the inequality $|x_j|\le(1+x_j^2)/2$ we conclude that the same is true
for odd $|\alpha|+|\beta|$.

Therefore, $\psi\in S_+$.

Theorem 1 is proved.

\subsection{Proof of Theorem 2}

Theorem 2 is deduced from Theorem 1 in the same way as Theorem 4 was deduced from Theorem 3. Q.~E.~D.

\section{Remarks}

1) Similar formulas, equations, and constructions appear in the theory of the Radon transform due to
I.~M.~Gelfand and others, cf. [1],~\S7 and references there.

2) A close description of the spinor representation of the orthogonal group is contained in the book
[12], Ch.~12.

3) The condition of being extended to the Siegel upper half-plane (and not only to the real Lagrangian Grassmannian)
in Theorem 1 is essential, as one can see already in the case
$n=1$. Indeed, consider a function $u(a)$ of one variable with compact support. If the function $u(a)$
cannot be extended to the upper half-plane, then its inverse Fourier transform is not supported on the positive
real axis, and it is easy to see that the corresponding function
$\psi(x)$ does not exist.

4) The question how to describe the Gaussian transform of the space $L_2$ seems not easy.
The construction of reproducing kernel (cf. [9], \S8) gives only a non-explicit description
of the Gaussian transform of the space $L_2$ as the image of an invariant integral operator on the Siegel upper
half-plane. This question is related to the following problem: what is the invariant formula
for the pairing between the spaces $S'$ and $S$ under the identifications from Theorems
1--4? In particular, this would give an invariant formula
for the $L_2$-norm, which could simplify the proof of Theorem 1. Formally it is not hard to write such a formula:
\begin{equation}
\const\cdot\int\limits_\SG u(Z)\overline{u(Z)}\frac{dZ\overline{dZ}}{(\det\Imm Z)^{n+\frac12}}.
\end{equation}
It is more difficult to give a sense to it. Seemingly, it should be done by analytical continuation
with respect to the exponent $n+\frac12$.
Cf. [11].

5) Under the identifications from Theorems 1--2,
the topology of the space $S$ goes to the $C^\infty$-topology on the space
of sections of a bundle on the Grassmannian $\Lambda_n$. This can be easily deduced from the proof of Theorems
1--2 (\S5). Convergence in this topology implies uniform convergence
of holomorphic functions on the Siegel upper half-plane and
$C^\infty$-convergence on any $\Sp(2n,\R)$-orbit in the closure of the upper half-plane.

6) The group $\Mp(2n,\R)$ acts also in the Gelfand--Shilov space $S_{1/2}^{1/2}(\R^n)$ [2]
and in the dual space $(S_{1/2}^{1/2})'(\R^n)$. One can ask whether the
Gaussian transform of the space
$S_{1/2}^{1/2}$ (respectively $(S_{1/2}^{1/2})'$) coincides with the space of analytical sections
of bundles $\mu^*$, $W\otimes\mu^*$ on the real Lagrangian Grassmannian, which satisfy equations (21),(26) and
can be analytically continued to the closure of the Siegel upper half-plane (respectively, with the space of
{\it all} holomorphic solutions of equations (17), (23) on the Siegel upper half-plane).
It is interesting whether there exist other functional spaces defined simply enough with an action of the
metaplectic group. It is known that the space $S$ coincides with the space of smooth vectors
in $L_2$ in the sense of representation theory of the group $\Mp(2n,\R)$ [9]. Does the space $S_{1/2}^{1/2}$
coincide with the (appropriately defined) space of analytical vectors in $L_2$?

7) {\it On infinite dimensional generalizations.} The expected infinite dimensional generalization of the
construction of the space $S_+$ from Theorem 1 can be performed in the following setup. Instead of the space
$\R^n$, one considers a real nuclear space $V$, say, the Schwartz space. Instead of the symplectic space $\R^{2n}$,
one considers the space $V\oplus V^*$ with its natural symplectic form.
Instead of the metaplectic group, one considers a central extension of the group of continuous symplectic automorphisms
of the space $V\oplus V^*$. The Lie algebra cocycle of this central extension is given by
\begin{equation}
\left(\left(\begin{array}{cc}a_1&b_1\\c_1&d_1\end{array}\right),
\left(\begin{array}{cc}a_2&b_2\\c_2&d_2\end{array}\right)\right)\mapsto\frac12\tr(b_1c_2-b_2c_1).
\end{equation}
Instead of determinant, one considers the Fredholm determinant (cf. [20]). Instead of symmetric (positive definite)
matrices, one considers symmetric continuous (positive definite) operators $V\to V^*$ or $V^*\to V$.

This way one can seemingly obtain a space analogous to $S_+$ and suitable for solution of equations of the form
\begin{equation}
\begin{aligned}{}
\frac{\partial\Psi}{\partial t}&=\int\left[a(x,y)u(x)\frac{\delta}{\delta u(y)}+\frac i2b(x,y)u(x)u(y)\right.\\
&\left.+\frac i2c(x,y)\frac{\delta}{\delta u(x)}\frac{\delta}{\delta u(y)}\right]\Psi\,dxdy,
\end{aligned}
\end{equation}
where $b(x,y)$ is a distribution, $c(x,y)$ is a smooth function, and $a(x,y)$ is the kernel of a linear operator
$V\to V$. But this equation does not look like equation (1), in which both $b$ and $c$ are singular distributions.
In addition, the evolution operators of the Klein--Gordon equation do not belong to this infinite dimensional
symplectic group. They are rather continuous operators $V\oplus V\to V\oplus V$, where $V$ is the Schwartz space.

Hence we come to a conclusion that this approach to equation (1) is not fruitful. For a more fruitful approach,
see [18].

\end{document}